\documentclass[twocolumn,aps,superscriptaddress,showpacs,floatfix,prc]{revtex4}
\usepackage{mathrsfs}
\usepackage{amssymb}
\usepackage{amsmath}
\usepackage{graphicx}
\usepackage[normalem]{ulem}
\usepackage[dvips]{color}
\usepackage{bm}
\usepackage{longtable}

\renewcommand\sout{\bgroup \color{red} \ULdepth=-.5ex \ULset}

\renewcommand{\v}[1]{\textbf{#1}}
\renewcommand{\rm}[1]{\textrm{#1}}
\renewcommand{\d}{\mathrm{d}}

\begin{document}

\title{The isospin quartic term in the kinetic energy of neutron-rich nucleonic matter}

\author{Bao-Jun Cai}
\affiliation{Department of Physics and Astronomy, Texas A$\&$M
University-Commerce, Commerce, TX 75429-3011, USA}
\author{Bao-An Li\footnote{%
Corresponding author: Bao-An.Li$@$tamuc.edu}}
\affiliation{Department of Physics and Astronomy, Texas A$\&$M
University-Commerce, Commerce, TX 75429-3011, USA}
\date{\today}

\begin{abstract}
The energy of a free gas of neutrons and protons is well known to be
approximately isospin parabolic with a negligibly small quartic term
of only $0.45$\,MeV at the saturation density of nuclear matter
$\rho_0=0.16\,\rm{fm}^{-3}$. Using an isospin-dependent
single-nucleon momentum distribution including a high (low) momentum
tail (depletion) with its shape parameters constrained by recent
high-energy electron scattering and medium-energy nuclear
photodisintegration experiments as well as the state-of-the-art
calculations of the deuteron wave function and the equation of state
of pure neutron matter near the unitary limit within several modern
microscopic many-body theories, we show for the first time that the
kinetic energy of interacting nucleons in neutron-rich nucleonic
matter has a significant quartic term of $7.18\pm2.52\,\rm{MeV}$.
Such a large quartic term has broad ramifications in determining the
equation of state of neutron-rich nucleonic matter using observables
of nuclear reactions and neutron stars.
\end{abstract}

 \maketitle

\section{Introduction} To determine the equation of state (EoS)
of isospin-asymmetric nuclear matter (ANM) has been a longstanding
goal shared by both nuclear physics and astrophysics\,\cite{EPJA}.
Usually one uses the so-called empirical parabolic law for the
energy per nucleon, i.e.,
$E(\rho,\delta)=E_0(\rho)+E_{\rm{sym}}(\rho)\delta^2+\mathcal{O}(\delta^4)$
where $\rho=\rho_{\rm{n}}+\rho_{\rm{p}}$ and
$\delta=(\rho_{\rm{n}}-\rho_{\rm{p}})/\rho$ are the nucleon density
and isospin asymmetry of the system in terms of the neutron and
proton densities $\rho_{\rm{n}}$ and $\rho_{\rm{p}}$, respectively.
The isospin quadratics of the ANM EoS has been verified to high
accuracies from symmetric ($\delta=0$) up to pure neutron
($\delta=1$) matter by most of the available nuclear many-body
theories using various interactions, see, e.g., ref.\,\cite{Bom91}.
Nevertheless, it has been shown consistently in a number of studies
that for some physical quantities relevant for understanding
properties of neutron stars, such as the proton fraction at $\beta$
equilibrium, core-crust transition density and the critical density
for the direct URCA process to happen, even a very small coefficient
$E_{\rm{sym,4}}(\rho)$ of the isospin quartic term in the EoS can
make a big difference \cite{Sjo74}.

Here we concentrate on examining the isospin quadratics of the
kinetic EoS. For many purposes in both nuclear physics and
astrophysics, such as simulating heavy-ion collisions\,\cite{LCK08}
and  determining critical formation densities of different charge
states of $\Delta$ resonances in neutron stars\,\cite{Cai14}, one
has to know separately the kinetic and potential parts of the EoS.
While neither any fundamental physical principle nor the empirical
parabolic law of the EoS requires the kinetic and potential parts of
the EoS to be quadratic in $\delta$ individually, in practice
especially in most phenomenological models the free Fermi gas (FFG)
EoS is often used for the kinetic part and then the generally less
known potential EoS is explored by comparing model predictions with
experimental data. It is well known that the FFG model predicts a
kinetic symmetry energy of $E^{\rm{kin}}_{\rm{sym}}(\rho_0)\approx
12.3$ MeV and a negligibly small quartic term of
$E^{\rm{kin}}_{\rm{sym},4}(\rho_0)=E^{\rm{kin}}_{\rm{sym}}(\rho_0)/27\approx
0.45$ MeV at $\rho_0=0.16/\rm{fm}^3$. However, nuclear interactions,
in particular the short-range repulsive core and tensor force, lead
to a high (low) momentum tail (depletion) in the single-nucleon
momentum distribution above (below) the nucleon Fermi
surface\,\cite{Mig57,bethe,pan92,Pan99}. Much progress has been made
recently both theoretically and experimentally in quantifying
especially the nucleon high momentum tails (HMT) in ANM, see, e.g.,
refs.\,\cite{Wei15,Wei15a,Hen15,Hen14,Egi06}. In this work, using
isospin-dependent nucleon HMT constrained by recent high-energy
electron scattering and medium-energy nuclear photodisintegration
experiments as well as the state-of-the-art calculations of the
deuteron wave function and the EoS of pure neutron matter (PNM) near
the unitary limit within several modern microscopic many-body
theories,  we show that the kinetic ANM EoS has a significant
quartic term of $E^{\rm{kin}}_{\rm{sym},4}(\rho_0)=7.18\pm
2.52\,\rm{MeV}$ that is about 16 times the FFG model prediction.

\section{Isospin dependence of single-nucleon momentum
distribution with a high momentum tail in neutron-rich matter}
Guided by well-known predictions of microscopic nuclear many-body
theories, see, e.g., reviews in ref.\,\cite{Ant88}, and recent
experimental findings\,\cite{Wei15,Wei15a,Hen15,Hen14}, we describe
the single-nucleon momentum distribution in ANM using
\begin{equation}\label{MDGen}
n^J_{\v{k}}(\rho,\delta)=\left\{\begin{array}{ll}
\Delta_J+\beta_J{I}\left(\displaystyle{|\v{k}|}/{k_{\rm{F}}^J}\right),~~&0<|\v{k}|<k_{\rm{F}}^J,\\
&\\
\displaystyle{C}_J\left({k_{\rm{F}}^{J}}/{|\v{k}|}\right)^4,~~&k_{\rm{F}}^J<|\v{k}|<\phi_Jk_{\rm{F}}^J.
\end{array}\right.
\end{equation}
Here, $J=\rm{n,p}$ is the isospin index,
$k_{\rm{F}}^J=k_{\rm{F}}(1+\tau_3^J\delta)^{1/3}$ is the transition
momentum\,\cite{Hen14} where $k_{\rm{F}}=(3\pi^2\rho/2)^{1/3}$ and
$\tau_3^{\rm{n}}=+1$, $\tau_3^{\rm{p}}=-1$.
\begin{figure}[h!]
\centering
  \includegraphics[width=8.cm]{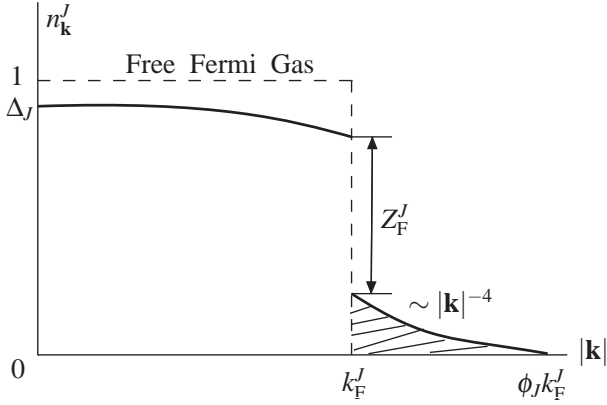}
  \caption{A sketch of the single-nucleon momentum distribution with a high momentum tail.}
  \label{mom-dis}
\end{figure}
The main features of $n^J_{\v{k}}(\rho,\delta)$ are depicted in Fig.
\ref{mom-dis}. The $\Delta_J$ measures the depletion of the Fermi sphere
at zero momentum with respect to the FFG model prediction while the
$\beta_J$ is the strength of the momentum dependence
$I(|\v{k}|/k_{\rm{F}}^J)$\,\cite{Bel61,Czy61,Sar80} of the depletion
near the Fermi surface. The jump $Z^J_{\rm{F}}$ of the momentum distribution at $k_{\rm{F}}^J$, namely, the ``renormalization
function",  contains information about the nucleon effective E-mass and its isospin dependence\cite{Jeu76}. Specifically,
$Z^J_{\rm{F}}=n_{k_{\rm{F}}^J-0}^J-n_{k_{\rm{F}}^J+0}^J={M}/{{M}_{\rm{E}}^{J,\ast}}$,
where $ {{M}^{J,\ast}_{\rm{E}}}/{M}\equiv[1-{\partial
V}/{\partial\omega}]^{-1}$ with $V$ and $\omega$ being the real part of the single-particle potential
and energy\,\cite{Mig57,Lut60}, respectively.

The amplitude ${C}_J$ and cutoff coefficient
$\phi_J$ determine the fraction of nucleons in the HMT via
\begin{equation}\label{xPNM}
x_J^{\rm{HMT}}=3C_{{J}}\left(1-\frac{1}{\phi_{{J}}}\right).
\end{equation}
The normalization condition $
[{2}/{(2\pi)^3}]\int_0^{\infty}n^J_{\v{k}}(\rho,\delta)\d\v{k}=\rho_J={(k_{\rm{F}}^{J})^3}/{3\pi^2}
$ requires that only three of the four parameters, i.e., $\beta_J$,
${C}_J$, $\phi_J$ and $\Delta_J$, are independent. Here we choose
the first three as independent and determine the $\Delta_J$ from
\begin{equation}\label{DeltaJ}
\Delta_J=1-\frac{3\beta_J}{(k_{\rm{F}}^{J})^3}\int_0^{k_{\rm{F}}^J}{I}\left(\frac{k}{k_{\rm{F}}^J}\right)k^2\d
k-3{C}_J\left(1-\frac{1}{\phi_J}\right).
\end{equation}

Hinted by the finding within the self-consistent Green function
(SCGF) theory\,\cite{Rio09} and the Brueckner-Hartree-Fock
(BHF) theory \,\cite{Yin13} the depletion $\Delta_J$ has an almost linear
dependence on $\delta$ in the opposite directions for neutrons and
protons, we expand all four parameters in the form
$Y_J=Y_0(1+Y_1^J\delta)$. Then, the total kinetic energy per nucleon
in ANM
\begin{equation}\label{kinE}
E^{\rm{kin}}(\rho,\delta)=\frac{1}{\rho}\frac{2}{(2\pi)^3}\sum_{J=\rm{n,p}}\int_0^{\phi_Jk_{\rm{F}}^J}\frac{\v{k}^2}{2M}n_{\v{k}}^J(\rho,\delta)\d\v{k}
\end{equation}
would obtain a linear term in $\delta$ of the form
\begin{align}\label{Ekin1}
&E_1^{\rm{kin}}(\rho)=\frac{3}{5}\frac{k_{\rm{F}}^2}{2M}\Bigg[\frac{5}{2}C_0\phi_0(\phi_1^{\rm{n}}+\phi_1^{\rm{p}})
\notag\\
&+\frac{5}{2}C_0(\phi_0-1)(C_1^{\rm{n}}+C_1^{\rm{p}})
+\frac{1}{2}\Delta_0(\Delta_1^{\rm{n}}+\Delta_1^{\rm{p}})\notag\\
&+\frac{5\beta_0(\beta_1^{\rm{n}}+\beta_1^{\rm{p}})}{2k_{\rm{F}}^5}
\int_0^{k_{\rm{F}}}I\left(\frac{k}{k_{\rm{F}}}\right)k^4\d k\Bigg]
\end{align}
where $M$ is the nucleon mass. To ensure that the
$E_1^{\rm{kin}}(\rho)$ vanishes as required by the neutron-proton
exchange symmetry of the EoS, we require that
$\Delta_1^{\rm{n}}=-\Delta_1^{\rm{p}}$,
$\beta_1^{\rm{n}}=-\beta_1^{\rm{p}}$,
${C}_1^{\rm{n}}=-{C}_1^{\rm{p}}$ and
$\phi_{1}^{\rm{n}}=-\phi_1^{\rm{p}}$, i.e., more compactly
$Y_J=Y_0(1+Y_1\tau_3^J\delta)$.

\section{Constraining the parameters of the single-nucleon
momentum distribution} It is well known that the nucleon HMT from
deuteron to infinite nuclear matter scales, see, e.g.,
refs.\,\cite{Fan84,Pie92,Cio96}, leading to constant per nucleon
inclusive $(\rm{e},\rm{e}^{\prime})$ cross sections for heavy nuclei
with respect to deuteron for the Bjorken scaling parameter
$x_{\rm{B}}$ between about 1.5 and 1.9, see, e.g.,
ref.\,\cite{Arr12} for a recent review. Systematic analyses of these
inclusive experiments and data from exclusive two-nucleon knockout
reactions induced by high-energy electrons or protons have firmly
established that the HMT fraction in symmetric nuclear matter (SNM)
is about $x^{\rm{HMT}}_{\rm{SNM}}=28\%\pm4$\% and that in PNM is
about
$x_{\rm{PNM}}^{\rm{HMT}}=1.5\%\pm0.5\%$\,\cite{Hen15,Hen14,Egi06,Hen14b}.

The ${C}/{|\mathbf{k}|^4}$ shape of the HMT for both SNM and PNM is
strongly supported by recent findings theoretically and
experimentally.  The HMT for deuteron from variational many-body
calculations using several modern nuclear forces decrease as
$|\mathbf{k}|^{-4}$ within about 10\% and in quantitative agreement
with that from analyzing the $\rm{d}(\rm{e}, \rm{e}^{\prime}\rm{p})$
cross section in directions where final state interaction suffered
by the knocked-out proton is small\,\cite{Hen15}. The extracted
magnitude $C_{\rm{SNM}}=C_0$ of the HMT in  SNM at $\rho_0$ is
${C}_0\approx 0.15\pm0.03$\,\cite{Hen15} (properly rescaled
considering the factor of 2 difference in the adopted normalizations
of $n_{\mathbf{k}}$ here and that in refs.\,\cite{Hen15,Hen14b}).
Rather remarkably, a very recent evaluation of medium-energy
photonuclear absorption cross sections has also presented clear and
independent evidence for the ${C}/{|\mathbf{k}|^4}$ behavior of the
HMT and extracted a value of ${C}_0\approx
0.172\pm0.007$\,\cite{Wei15} for SNM at $\rho_0$ in very good
agreement with that found in ref.\,\cite{Hen15}. In the following,
we use $C_0\approx0.161\pm0.015$ from taking the average of the
above two constraints. With this $C_0$ and the value of
$x^{\rm{HMT}}_{\rm{SNM}}$ given earlier, the HMT cutoff parameter in
SNM is determined to be
$\phi_0=(1-x_{\rm{SNM}}^{\rm{HMT}}/3{C}_0)^{-1}=2.38\pm0.56$.

Very interestingly, the ${1}/{|\mathbf{k}|^4}$ behavior of the HMT
nucleons is identical to that in two-component (spin-up and -down)
cold fermionic atoms first predicted by Tan\,\cite{Tan08} and then
quickly verified experimentally\,\cite{Ste10}. Tan's general
prediction is for all two-component fermion systems having an s-wave
contact interaction with a scattering length $a$ much larger than
the inter-particle distance $d$ which has to be much longer than the
interaction range $r_{\rm{e}}$. At the unitary limit when
$|k_{\rm{F}}a|\rightarrow \infty$, Tan's prediction is universal for
all fermion systems. Since the HMT in nuclei and SNM is known to be
dominated by the tensor force induced neutron-proton pairs with the
$a\approx 5.4$ fm and $d\approx 1.8$ fm at $\rho_0$, as noted in
refs.\,\cite{Hen15,Wei15}, Tan's stringent conditions for unitary
fermions is obviously not satisfied in normal nuclei and SNM. The
observed identical ${1}/{|\mathbf{k}|^4}$ behavior of the HMT in
nuclei and cold atoms may have some deeper physical reasons
deserving further investigations. Indeed, a very recent study on the
$A(\rm{e,e}'\rm{p})$ and $A(\rm{e,e}'\rm{pp})$ scattering has shown
that the majority of the short range correlation (SRC)-susceptible
n-p pairs are in the $^3\rm{S}_1$ state\,\cite{Col15}. On the other
hand, because of the unnaturally large neutron-neutron scattering
length $a_{\rm{nn}}(^1\rm{S}_0)=-18.8$ fm, it is known that PNM is
closer to the unitary limit\,\cite{Sch05}. The EoS of PNM can thus
be expanded as\,\cite{Bak99}
\begin{equation}\label{EPNMTan}
E_{\rm{PNM}}(\rho)\simeq\frac{3}{5}\frac{(k^{\textrm{PNM}}_{\textrm{F}})^2}{2M}\left[\xi-\frac{\zeta}{k^{\textrm{PNM}}_{\textrm{F}}a_{\textrm{nn}}}
-\frac{5\nu}{3(k^{\textrm{PNM}}_{\textrm{F}}a_{\textrm{nn}})^2}\right],
\end{equation}
where $k_{\rm{F}}^{\rm{PNM}}=2^{1/3}k_{\rm{F}}$ is the transition
momentum in PNM, $\xi\approx0.4\pm0.1$ is the Bertsch
parameter\,\cite{BertPara}, $\zeta\approx\nu\approx1$ are two
universal constants\,\cite{Bul05}.

\begin{figure}[h!]
\centering
\includegraphics[width=8.5cm]{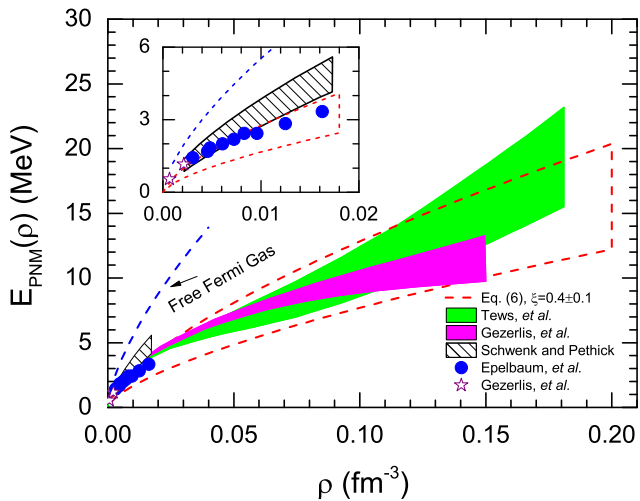}
\caption{(Color Online) The EoS of PNM obtained from Eq.
(\ref{EPNMTan}) (dashed red band) and that from next-leading-order
(NLO) lattice calculation\,\cite{Epe09a} (blue solid points), chiral
perturbative theories\,\cite{Tew13} (green band), quantum Monte
Carlo simulations (QMC)\,\cite{Gez13,Gez10} (magenta band and purple
stars), and effective field theory\,\cite{Sch05}.}\label{TanPNMLow}
\end{figure}

Shown in Fig. \ref{TanPNMLow} is a comparison of the EoS of PNM
obtained from Eq. (\ref{EPNMTan}) (dashed red band) with several
state-of-the-art calculations using modern microscopic many-body
theories. At densities less than about 0.01\,$\textrm{fm}^{-3}$, as
shown in the inset, the Eq. (\ref{EPNMTan}) is consistent with the
prediction by the effective field theory\,\cite{Sch05}. In the range
of 0.01\,$\textrm{fm}^{-3}$ to about $0.02\,\textrm{fm}^{-3}$, it
has some deviations from predictions in ref.\,\cite{Sch05} but
agrees very well with the NLO lattice simulations\,\cite{Epe09a}. At
higher densities up to about $\rho_0$, it overlaps largely with
predictions by the chiral perturbation theories\,\cite{Tew13} and
the quantum Monte Carlo simulations\,\cite{Gez13,Gez10}.
In addition, recent studies on the spin-polarized neutron matter within
the chiral effective field theory including two-, three-,
and four-neutron interactions indicate that properties of PNM
is similar to the unitary Fermi gas at least upto $\rho_0$  far beyond the scattering-length regime
of $\rho\lesssim\rho_0/100$\,\cite{Kru15}.
Overall, the above comparison and studies clearly justify the use of Eq. (\ref{EPNMTan}) to
calculate the PNM EoS up to about $\rho_0$.

Both the HMT and EoS can be experimentally measured independently and calculated simultaneously within the same model.
Tan has proven in great detail that the two are  directly related by the so-called adiabatic sweep theorem \cite{Tan08}.
It is valid for any two-component Fermi systems under the same conditions as the Eq. \ref{EPNMTan} near the unitary limit. For PNM, it can be written as
\begin{equation}\label{ast}
C_{\rm{n}}^{\rm{PNM}}\cdot(k_{\rm{F}}^{\rm{PNM}})^4=-4\pi
M\cdot\frac{\d (\rho E_{\rm{PNM}})}{\d(a^{-1})}.
\end{equation}
While the results shown in Fig. \ref{TanPNMLow} justify the use of Eq. \ref{EPNMTan} for the EoS of PNM up to about $\rho_0$,
indeed, to our best knowledge there is currently no proof that the Eq. \ref{ast} is also valid in the same density range as the Eq. \ref{EPNMTan}.
Thus, it would be very interesting to examine the validity range of Eq. \ref{ast} using the same models as those used to calculate the EoS. In this work, we assume
that the Eqs.  \ref{EPNMTan} and \ref{ast} are both valid in the same density range. Then, the strength of the HMT in PNM can be readily obtained as
\begin{equation}
C_{\rm{n}}^{\rm{PNM}}\approx 2\zeta/5\pi+4\nu/(3\pi
k_{\rm{F}}^{\rm{PNM}}a_{\rm{nn}}(^1\rm{S}_0))\approx0.12.
\end{equation}
Noticing that $C_{\rm{n}}^{\rm{PNM}}=C_0(1+C_1)$, we can then infer that
$C_1=-0.25\pm0.07$ with the $C_0$ given earlier. Next, after
inserting the values of $x_{\rm{PNM}}^{\rm{HMT}}$ and
$C_{\rm{n}}^{\rm{PNM}}$ into Eq. ({\ref{xPNM}), the high momentum
cutoff parameter for PNM is determined to be
$\phi_{\rm{n}}^{\rm{PNM}}\equiv
\phi_0(1+\phi_1)=(1-x_{\rm{PNM}}^{\rm{HMT}}/3C_{\rm{n}}^{\rm{PNM}})^{-1}=1.04\pm0.02$.
It is not surprising that the $\phi_{\rm{n}}^{\rm{PNM}}$ is very close to
unity since only about 1.5\% neutrons are in the HMT in
PNM. Subsequently, using the $\phi_0$ determined earlier, we get
$\phi_1=-0.56\pm0.10$.

The two parameters $\beta_0$ and $\beta_1$  in
$\beta_J=\beta_0(1+\beta_1\tau_3^J\delta)$ depend on the function
$I(|\v{k}|/k_{\rm{F}}^J)$ which is still model dependent. To
minimize the model assumptions and evaluate the dominating terms in
the kinetic EoS, in the following we shall first use a
momentum-independent depletion of the Fermi sea as in most studies
in the literature. The HMT parameters $C_J$ and $\phi_J$  evaluated
above remain the same. Then,  we examine the maximum correction to
each term in the kinetic EoS by using  the largest values of
$\beta_0$ and $\beta_1$ allowed and a typical function
$I(|\v{k}|/k_{\rm{F}}^J)$. Not surprisingly, the corrections are all
small.

\section{Isospin dependence of kinetic EoS of ANM} The kinetic
EoS can be expanded in $\delta$ as
\begin{equation}
E^{\rm{kin}}(\rho,\delta)=E_0^{\rm{kin}}(\rho)+E_{\rm{sym}}^{\rm{kin}}(\rho)\delta^2+E_{\rm{sym,4}}^{\rm{kin}}(\rho)\delta^4+\mathcal{O}(\delta^6).
\end{equation}
The coefficients evaluated from Eq. (\ref{kinE}) using the $n^J_{\v{k}}(\rho,\delta)$ in Eq. (\ref{MDGen}) with $\beta_J=0$ are
\begin{align}
E^{\rm{kin}}_0(\rho)=&\frac{3}{5}E_{\rm{F}}(\rho)\left[
1+{C}_0\left(5\phi_0+\frac{3}{\phi_0}-8\right)\right],\label{E0kin}\\
E_{\rm{sym}}^{\rm{kin}}(\rho)=&\frac{1}{3}E_{\rm{F}}(\rho)\Bigg[1+{C}_0\left(1+3{C}_1\right)\left(5\phi_0+\frac{3}{\phi_0}-8\right)\notag\\
&\hspace*{-1.5cm}+3{C}_0\phi_1\left(1+\frac{3}{5}{C}_1\right)\left(5\phi_0-\frac{3}{\phi_0}\right)+\frac{27{C}_0\phi_1^2}{5\phi_0}\Bigg],\label{Esymkin}\\
E_{\rm{sym,4}}^{\rm{kin}}(\rho)=&\frac{1}{81}E_{\rm{F}}(\rho)\Bigg[1+{C}_0(1-3{C}_1)\left(5\phi_0+\frac{3}{\phi_0}
-8\right)\notag\\
&\hspace*{-1.5cm}+3{C}_0\phi_1(9{C}_1-1)\left(5\phi_0-\frac{3}{\phi_0}\right)\notag\\
&\hspace*{-1.5cm}+\frac{81{C}_0\phi_1^2(9\phi_1^2-9{C}_1\phi_1-15\phi_1+15{C}_1+5)}{5\phi_0}
\Bigg].\label{Esymkin4}
\end{align}
In the FFG where there is no HMT, $\phi_0=1$, $\phi_1=0$ and thus
$5\phi_0+3/\phi_0-8=0$, the above expressions reduce naturally to the well
known results of $E^{\rm{kin}}_0(\rho)=3E_{\rm{F}}(\rho)/5$,
$E_{\rm{sym}}^{\rm{kin}}(\rho)=E_{\rm{F}}(\rho)/3$, and
$E_{\rm{sym,4}}^{\rm{kin}}(\rho)/E_{\rm{sym}}^{\rm{kin}}(\rho)=1/27$
where $E_{\rm{F}}(\rho)=k_{\rm{F}}^2/2M$ is the Fermi energy.

For the interacting nucleons in ANM with the momentum distribution and its parameters given earlier, we found that
$E_0^{\rm{kin}}(\rho_0)=40.45\pm8.15\,\rm{MeV}$,
$E_{\rm{sym}}^{\rm{kin}}(\rho_0)=-13.90\pm11.54\,\rm{MeV}$ and
$E_{\rm{sym,4}}^{\rm{kin}}(\rho_0)=7.19\pm2.52\,\rm{MeV}$,
respectively. Compared to the corresponding values for the FFG, it
is seen that the isospin-dependent HMT increases significantly the average kinetic
energy $E_0^{\rm{kin}}(\rho_0)$ of SNM but decreases the kinetic
symmetry energy $E_{\rm{sym}}^{\rm{kin}}(\rho_0)$ of ANM to a negative value qualitatively consistent with
findings of several recent studies of the kinetic EoS considering short-range nucleon-nucleon correlations using both phenomenological
models and microscopic many-body theories\,\cite{CXu11,Vid11,Lov11,Car12,Rio14,Car14}.
However, it was completely unknown before if the empirical isospin parabolic law is still valid for the kinetic EoS of ANM
when the isospin-dependent HMTs are considered. Very surprisingly and interestingly, our calculations here show clearly
that it is broken seriously. More quantitatively, the ratio
$|E_{\rm{sym,4}}^{\rm{kin}}(\rho_0)/E_{\rm{sym}}^{\rm{kin}}(\rho_0)|$
is about $52\%\pm26$\% that is much larger than the FFG value of
$3.7$\%. We also found that the large quartic term is mainly due to the
isospin dependence of the HMT cutoff described by the $\phi_1$ parameter.
For example, by artificially setting $\phi_1=0$, we obtain
$E^{\rm{kin}}_{\rm{sym}}(\rho_0)=14.68\pm2.80$ MeV and
$E^{\rm{kin}}_{\rm{sym},4}(\rho_0)=1.12\pm0.27$ MeV which are all
close to their FFG values.

Considering short-range nucleon-nucleon correlations but assuming
that the isospin parabolic approximation is still valid, some
previous studies have evaluated the kinetic symmetry energy
$E_{\rm{sym}}^{\rm{kin}}$ by taking the difference between the
kinetic energies of PNM and SNM, i.e., subtracting the
$E_{\rm{PNM}}^{\rm{kin}}$ by $E_0^{\rm{kin}}$. This actually
approximately equals to
$E_{\rm{sym}}^{\rm{kin}}(\rho_0)+E_{\rm{sym,4}}^{\rm{kin}}(\rho_0)=-6.71\pm9.11$\,MeV
in our current work. This value is consistent quantitatively with
the $E_{\rm{sym}}^{\rm{kin}}(\rho_0)$ found in ref.\,\cite{Hen14b}
using the parabolic approximation.

\section{Corrections due to the momentum-dependent depletion of
the Fermi sea} To estimate corrections due to the momentum
dependence of the depletion very close to the Fermi surface, i.e., a
finite $\beta_J$, we consider a widely used single-nucleon momentum
distribution parameterized in ref.\,\cite{Cio96} based on
calculations using many-body theories. For
$|\v{k}|\lesssim2\,\rm{fm}^{-1}$, it goes like $\sim
e^{-\alpha|\v{k}|^2}$ with $\alpha\approx 0.12\,\rm{fm}^2$. At
$\rho_0$ since $\alpha k_{\rm{F}}^2\approx0.21$,
$e^{-\alpha|\v{k}|^2}\approx1-\alpha|\v{k}|^2+\mathcal{O}(|\v{k}|^4)$
is a good approximation in the range of $0<|\v{k}|<k_{\rm{F}}^J$.
Thus, we adopt a quadratic function
${I}(|\v{k}|/k_{\rm{F}}^J)=(|\v{k}|/k_{\rm{F}}^J)^2$. The constants
in the parameterization of ref.\,\cite{Cio96} are absorbed into our
parameters $\Delta_J$ and $\beta_J$. Then Eq. (\ref{DeltaJ}) gives
us $\Delta_J=1-3\beta_J/5-3{C}_J\left(1-1/\phi_J\right)$.
Specifically, we have $\beta_0=(5/3)[1-\Delta_0-3C_0(1-\phi_0^{-1})]=(5/3)[1-\Delta_0-x_{\rm{SNM}}^{\rm{HMT}}]$
for SNM. Then using the predicted value  of $\Delta_0\approx 0.88\pm0.03$\,\cite{Yin13,Pan99,Fan84} and the experimental value
of $x_{\rm{SNM}}^{\rm{HMT}}\approx 0.28\pm 0.04$, the value of $\beta_0$ is estimated to be about $-0.27\pm0.08$.
Similarly, the condition $\beta_J=\beta_0(1+\beta_1\tau_3^J\delta)<0$, i.e., $n_{\v{k}}^J$ is
a decreasing function of momentum towards $k_{\rm{F}}^J$, indicates that $|\beta_1|\leq1$.

\begin{figure}[h!]
\centering
  \includegraphics[width=7.5cm]{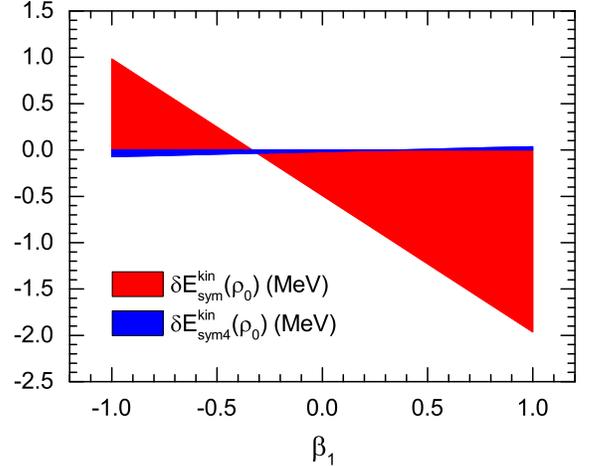}
  \caption{(Color Online) Corrections to the $E_{\rm{sym}}^{\rm{kin}}(\rho_0)$ and $E_{\rm{sym,4}}^{\rm{kin}}(\rho_0)
  $ as functions of $\beta_1$ with $\beta_0=-0.35$.}
  \label{fig_betaJeffect}
\end{figure}
First of all, a finite value of $\beta_J$ is expected to affect the
``renormalization function"  $Z^J_{\rm{F}}$. For SNM, we have
$Z_{\rm{F}}^0=1+2\beta_0/5-C_0-x_{\rm{SNM}}^{\rm{HMT}}= 0.45\pm0.07$
($0.56\pm0.04$) in the presence (absence) of $\beta_0$. For ANM,
however, the $Z^J_{\rm{F}}$ depends on the less constrained value of
$\beta_1$. It is worth noting that the latter also determines the
neutron-proton effective E-mass splitting which has significant
effects on isovector observables in heavy-ion
collisions\,\cite{BALi15}, and a study is underway to further
constrain the value of $\beta_1$ using data from heavy-ion
reactions.

Contributions from a finite $\beta_J$ to the first three terms of
the kinetic EoS are
\begin{align}
\delta
E_{0}^{\rm{kin}}(\rho)=&\frac{3}{5}E_{\rm{F}}(\rho_0)\cdot\frac{4\beta_0}{35},\\
~~\delta
E_{\rm{sym}}^{\rm{kin}}(\rho)=&\frac{1}{3}E_{\rm{F}}(\rho_0)\cdot\frac{4\beta_0(1+3\beta_1)}{35},\\
\delta
E_{\rm{sym,4}}^{\rm{kin}}(\rho)=&\frac{1}{81}E_{\rm{F}}(\rho_0)\cdot\frac{4\beta_0(1-3\beta_1)}{35}.
\end{align}
With the largest magnitude of $\beta_0=-0.35$, we examine  in Fig. \ref{fig_betaJeffect}
the corrections to the $E_{\rm{sym}}^{\rm{kin}}(\rho_0)$ and
$E_{\rm{sym,4}}^{\rm{kin}}(\rho_0)$ as functions of $\beta_1$ in its
full range allowed. In this case the maximum effects of the finite
$\beta_J$ are revealed. It is seen that the correction on the
$E_{\rm{sym,4}}^{\rm{kin}}(\rho_0)$ is negligible while the
correction on the $E_{\rm{sym}}^{\rm{kin}}(\rho_0)$ is less than
2\,MeV. Considering the corrections due to the finite
$\beta_0$ and $\beta_1$ and their uncertainties, we finally obtain
$E_0^{\rm{kin}}(\rho_0)=39.77\pm8.13\,\rm{MeV}$,
$E_{\rm{sym}}^{\rm{kin}}(\rho_0)=-14.28\pm11.59\,\rm{MeV}$ and
$E_{\rm{sym,4}}^{\rm{kin}}(\rho_0)=7.18\pm2.52\,\rm{MeV}$,
respectively. We notice here that the $\delta^6$ term was also consistently
evaluated and was found to be negligibly small at $\rho_0$.

\section{Summary and Discussions} In summary, using an isospin-dependent
single-nucleon momentum distribution including a high (low) momentum
tail (depletion) with its shape parameters constrained by the latest
results of several relevant experiments and the state-of-the-art
predictions of modern microscopic many-body theories, we found for
the first time that the kinetic EoS of interacting nucleons in ANM
is not parabolic in isospin asymmetry. It has a significant quartic
term of $7.18\pm2.52\,\rm{MeV}$ while its quadratic term is
$-14.28\pm11.60\,\rm{MeV}$ at saturation density of nuclear matter.

To this end, it is necessary to point out the limitations of our
approach and a few physical implications of our findings. Since we
fixed the parameters of the nucleon momentum distribution (Eq.
(\ref{MDGen})) by using experimental data and/or model calculations
at the saturation density, the possible density dependence of these
parameters is not explored in this work. The density dependence of
the various terms in the kinetic EoS is thus only due to that of the
Fermi energy as shown in Eqs.(\ref{E0kin})-(\ref{Esymkin4}). In this
limiting case, the slope of the kinetic symmetry energy, i.e.,
$L^{\rm{kin}}=3\rho_0\partial
E_{\rm{sym}}^{\rm{kin}}(\rho)/\partial\rho|_{\rho=\rho_0}=-27.81\pm
23.08$ MeV while that of the FFG is about 25.04 MeV.

The SRC-reduced kinetic symmetry energy with respect to the FFG
prediction has been found to affect significantly not only our
understanding about the origin of the symmetry energy but also
several isovector observables, such as the free neutron/proton and
$\pi^-/\pi^+$ ratios in heavy-ion collisions
\cite{Hen14b,Li15,Yong15}.  However, to our best knowledge, an
investigation on possible effects of a large isospin quartic term on
heavy-ion collisions has never been done while its effects on
properties of neutron stars have been studied extensively
\cite{Sjo74}. Of course, effects of the quartic and quadratic terms
should be studied together within the same approach. To extract from
nuclear reactions and neutron stars information about the EoS of
neutron-rich matter, people often parameterize the EoS as a sum of
the kinetic energy of a FFG and a potential energy involving unknown
parameters upto the isospin-quadratic term only.  Our findings in
this work indicate that it is important to include the
isospin-quartic term in both the kinetic and potential parts of the
EoS. Moreover, to accurately extract the completely unknown
isospin-quartic term $E^{\rm{pot}}_{\rm{sym,4}}(\rho)\delta^4$ in
the potential EoS it is important to use the kinetic EoS of
quasi-particles with reduced kinetic symmetry energy and an enhanced
quartic term due to the isospin-dependence of the HMT. Most relevant
to the isovector observables in heavy-ion collisions, such as the
neutron-proton ratio and differential flow, is the nucleon isovector
potential. Besides the so-called Lane potential $\pm 2\rho
E^{\rm{pot}}_{\rm{sym}}(\rho)\delta$ where the
$E^{\rm{pot}}_{\rm{sym}}(\rho)$ is the potential part of the
symmetry energy and the $\pm$ sign is for neutrons/protons, the
$E^{\rm{pot}}_{\rm{sym,4}}(\rho)\delta^4$ term contributes an
additional isovector potential  $\pm 4 \rho
E^{\rm{pot}}_{\rm{sym,4}}(\rho)\delta^3$. In neutron-rich systems
besides neutron stars, such as nuclear reactions induced by rare
isotopes and peripheral collisions between two heavy nuclei having
thick neutron-skins, the latter may play a significant role in
understanding the isovector observables or extracting the sizes of
neutron-skins of the nuclei involved. We plan to study effects of
the isospin-quartic term in the EoS in heavy-ion collisions using
the isospin-dependent transport model \cite{LCK08} in the near
future.

\section{Acknowledgement} We would like to thank L.W. Chen, O. Hen, X.H. Li, W.G.
Newton, E. Piasetzky, A. Rios, I. Vida$\tilde{\textrm{n}}$a and L.B.
Weinstein for helpful discussions. This work is supported in part by
the US National Science Foundation under Grant No. PHY-1068022 and the U.S. Department of Energy Office of Science under Award Number DE-SC0013702.

\end{document}